\documentclass{PoS}
\usepackage{amsmath}
\usepackage{enumerate}

\title{	Prospects for the Use of Photosensor Timing Information with Machine Learning Techniques in Background Rejection}

\ShortTitle{Photosensor Timing Information with Machine Learning }

\author{\speaker{S.~Spencer} $^a$,
T.~Armstrong$\,^a$,
J.~Watson$\,^{a,b}$,
G.~Cotter$\,^a$
\\
\llap{$^a$} Department of Physics, University of Oxford, Keble Road, Oxford, OX1 3RH, United Kingdom \\
\llap{$^b$} Max-Planck-Institut f{\"u}r Kernphysik, P.O. Box 103980, 69029 Heidelberg, Germany \\

Email: \email{samuel.spencer@physics.ox.ac.uk}
}

\abstract{
Recent developments in machine learning (ML) techniques present a promising new analysis method for high-speed imaging in astroparticle physics experiments, for example with imaging atmospheric Cherenkov telescopes (IACTs). In particular, the use of timing information with new machine learning techniques provides a novel method for event classification. Previous work in this field has utilised images of the integrated charge from IACT camera photomultipliers, but the majority of current and upcoming IACT cameras have the capacity to read out the entire photosensor waveform following a trigger. As the arrival times of Cherenkov photons from extensive air showers (EAS) at the camera plane are dependent upon the altitude of their emission, these waveforms contain information useful for IACT event classification. In this work, we investigate the potential for using these waveforms with ML techniques, and find that a highly effective means of utilising their information is to create a set of seven additional two dimensional histograms of waveform parameters to be fed into the machine learning algorithm along with the integrated charge image. This appears to be superior to using only these new ML techniques with the waveform integrated charge alone. We also examine these timing-based ML techniques in the context of other experiments.
}

\FullConference{36th International Cosmic Ray Conference -ICRC2019-\\
		July 24th - August 1st, 2019\\
		Madison, WI, U.S.A.}
	
\begin{document}
\section{Introduction}
Imaging Atmospheric Cherenkov Telescopes (IACTs) are a class of indirect detectors of astrophysical gamma-rays. They operate by detecting Cherenkov light emitted from a particle cascade or Extensive Air Shower (EAS) created when a high energy gamma-ray hits the top of Earth's atmosphere. In order to better reconstruct the point of origin of the gamma-ray, most IACTs operate stereoscopically in arrays. There are currently three major IACT arrays: H.E.S.S. in Namibia, VERITAS in Arizona and MAGIC on La Palma \cite{scienceCTA}. . The  Cherenkov  Telescope  Array  (CTA)  is  an  ambitious  project  to  build  a  next-generation IACT array, which would improve on the current generation instruments' angular resolution and energy  sensitivity  by  roughly  an  order  of  magnitude \cite{scienceCTA}. 

High energy cosmic ray protons and electrons are a significant background for IACTs, and classifying events as such is a major limiting factor in the sensitivity of the instrument. In the case of protons, because their quarks experience the strong nuclear force, the typical interactions they undergo on entering the atmosphere are different to photons. This results in a wider overall EAS compared to a gamma-ray induced shower \cite{hillas}, as well as electromagnetic substructure to the shower created by the production of high-energy gamma-rays.  In the case of cosmic ray electrons, they are difficult to distinguish from gamma-rays as they undergo similar interactions. The only significant difference between the two is the altitude at which they typically first interact \cite{electronref}.

The basis of the methods for event classification in the current generation of IACTs are Hillas Parameters. They are obtained \cite{hillas} from the second moments of an IACT camera image (constructed from the integrated charge for each photomultiplier pixel). In order to take into account information from all of the telescopes in the array, these parameters can be combined into the Mean Reduced Scaled Width and Length (MRSW/MRSL) \cite{hessbdt}. Initially Hillas Parameters were simply used to perform simple data cuts to separate hadronic showers from gamma-ray induced showers based on their differing morphology. However, in recent years, more sophisticated methods using Boosted Decision Trees (BDTs) (taking the MRSL, MRSW, integrated charge and other parameters derived from Monte-Carlo lookup tables) eventually became the preferred methods for incident particle classification \cite{hessbdt}. Additionally for point source analyses, it is common to perform cuts on $\theta^2$ (the square of the angular difference between the reconstructed shower position and the source position) \cite{hessbdt}. 

The disadvantage of these techniques is that they don't take advantage of the full camera image of the EAS. As such, subtle details in the images, which may provide useful information, are not taken into account. This becomes an issue at the high and low energy boundaries that CTA is hoping to break, as hadronic and electron induced showers can closely resemble gamma-ray induced ones. This motivates us to investigate new analysis techniques for event discrimination in order to effectively improve our IACT sensitivity \cite{sitlowe}. 

\section{Previous Work}
Shilon et.al. in 2018 applied modern convolutional neural network based background rejection methods to H.E.S.S. data \cite{Shilon}. In their work, they feed images from the CT1-4 telescopes into convolutional layers that then feed into an Long Short-Term Memory Network (LSTM). This technique, called the CRNN method, treats the set of images from the four IACTs as a time series. Using this, they observed performance in background rejection superior to the current BDT paradigm with simulations, and were able to perform a significant detection of a flare in PKS-2155 with real data. 

Whilst work on the use of EAS timing information for event classification in IACTs has been investigated before \cite{Paulathesis}, the work in these proceedings is the first use of timing information in combination with modern ML techniques. This is not however the case for other air shower experiments, and in this section we investigate the approaches taken by other groups.

Erdmann et.al. \cite{aug1} investigated the use of deep learning to perform energy and angular event reconstruction of air showers for a model hexagonal array of 81 ground based water Cherenkov detectors using a custom simulation code. They feed in waveforms from their array into a CNN, which is concatenated with 2D histograms of the first particle arrival time and integrated charge for the entire detector grid approximately half way through their network architecture. These feature maps are then treated with separable convolutions, as it is not expected that the correlations between the feature maps will be coupled to spatial correlations within the maps themselves. The use of this data allows them to achieve an energy resolution of 5\% in the EeV energy range, compared to 8\% without the additional concatenated histograms.

In \cite{icecube1}, the IceCube collaboration perform reconstruction of Muon-Neutrino events using deep learning. Each of their 5160 digital optical modules (DOMs) records a waveform when Cerenkov light emitted from a particle interaction in the ice reaches the photomultiplier in the DOM. In that work, they parameterize their waveforms into 7 variables which are then inserted into a 3D neural network. On simulated events, they are able to obtain an improvement of almost 50\% in energy resolution relative to a more conventional reconstruction method. 

\section{Methodology}
\subsection{Simulations and Data Pre-processing}
Many of the camera prototypes under development for CTA have the ability to read out the entirety of their photomultiplier waveforms. One of our main goals is to find a way to use this extra information in combination with deep learning techniques in order to take maximal advantage of the information available about the EAS as possible. For example, using the time associated with the peak of the waveform for each pixel in the camera may aid in determining the primary particle's characteristics. 

To investigate this, we require a dataset of labelled simulations of gamma-ray, hadronic and electron induced EAS as they would appear to an IACT array. To generate this, we used the Corsika version 6.990 and Sim\_telarray version 2017-09-01 monte-carlo simulation packages \cite{BERNLOHR}. Parameters describing the Corsika simulations are shown in Table 1. The simulated array was of 4 Gamma Cherenkov Telescopes (GCTs) \cite{gct} equipped with the current version of the Compact High Energy Cameras (CHEC) \cite{checmpaper}, although the conclusions of this work are applicable to other telescopes with similar capabilities. The telescopes are arranged in a cross-configuration (+) on the Paranal site separated from the centre of the array by +-80 meters. This array configuration was chosen both to resemble the existing literature \cite{Shilon} and produce data that could be processed within our constraints of GPU resources. 

We investigate two scenarios. In the first, we attempt to differentiate between gamma-rays from a point source against a diffuse background of proton and electron events (as would be the case for most observations). In the second, we attempt to differentiate between diffuse gamma-ray events, diffuse proton events and diffuse electron events.

The data from the event files generated by Corsika/sim\_telarray were then fed into a python script to calibrate them using the prototype CTApipe (v0.6.1) software package \cite{ctapipe} and extract the waveform parameters. The gamma-ray, proton and electron data is then randomly mixed together in equal quantities and then saved in the HDF5 file format for convenience and storage.

\begin{table}[t]
\centering
\begin{tabular}{r|c|c|c|c}
    \textbf{Parameter} &\textbf{Point Source $\gamma$} & \textbf{Diffuse $\gamma$} & \textbf{Diffuse p} &\textbf{Diffuse e}\\
    \hline
    No. Training Events Point Source Run& 360995 &-&361135 &361135\\
    No. Testing Events Point Source Run& 239128&-&239227&239227\\
    No. Training Events Diffuse Run& - &365763&365903&365903\\
    No. Testing Events Diffuse Run& -&240453&240552&239894\\
    Energy Range (TeV) &0.3-330&0.3-330&1-600&0.3-330\\
    View Cone &$0^{\circ}$&$0^{\circ}$-$10^{\circ}$&$0^{\circ}$-$10^{\circ}$&$0^{\circ}$-$10^{\circ}$\\
\end{tabular}
\caption{Dataset Parameters Used. All simulation runs had a spectral index of -2, a zenith angle of $20^{\circ}$, an azimuth angle of $0^{\circ}$ and a Cherenkov emission waveband of 240-700nm.}
\end{table}
\subsection{Machine Learning Process}
\begin{figure}[t]
  \centering
  \includegraphics[width=\textwidth]{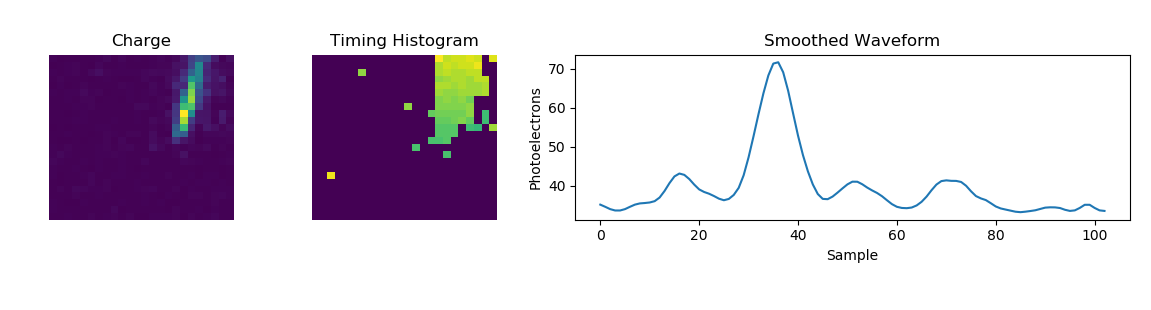}
  \vspace{-1.2cm}
  \caption{Left: The charge image from a 44TeV proton event. Centre: The associated timing histogram. Right: The first smoothed waveform to survive the peak charge cut in the image.}
  \label{fig:wfplot}
\end{figure}
In order to perform our analysis, we used the \textit{Keras} \cite{Keras} Python package with the \textit{Tensorflow} backend. We also used the \textit{scikit-learn} \cite{scikit} package to analyse the results. In contrast to the Shilon et.al. paper, we do not use their CRNN architecture, opting for a ConvLSTM2D approach instead \cite{shi}. However, the principle of using a mixture of convolutional layers and recurrence to handle data from multiple telescopes is the same. We compared four different techniques for classifying events on a like for like basis based on these networks. In the first (Method A), we use the available waveform information to generate a 2D timing histogram across the camera plane. We then perform a cut on the peak amplitude of the waveforms at 70 photoelectrons and set the other pixels in the histogram to zero in order to highlight the position of the shower in the image. These four timing histograms (one for each telescope) are then fed into the network without anything else, in the order of increasing median photon arrival time. In the second case (Method B), we feed in the integrated charge image as the first channel to the ConvLSTM2D. We then feed in the timing histogram created using the same technique as Method A as the second channel, and similarly cut 2D histograms of six other parameters from the smoothed waveform: the mean amplitude, the peak amplitude, the root mean square value, the full width at half maximum, the pulse rise time (RT) and the pulse fall time (FT). The third (Method C) uses integrated charge as a time proxy, as in the Shilon et.al. paper, and nothing else. The four images are then sorted and fed into the ConvLSTM2D such that the images with the largest integrated charge are fed in first. In the final method (Method D), we extract the median photon peak time from the waveforms and use this in place of using integrated charge as a time proxy. In this case the only images actually fed to the network are of integrated charge. As such the only difference between methods C and D is the order in which the images are fed into the ConvLSTM2D.

The datasets used, along with all training hyperparameters, was the same across all the four methods. As this is a relative study, we did not perform hyperparameter optimisation. The ConvLSTM2D network architecture was also identical across the four methods, with the exception of the shape of the input vector. The network architecture we use consists of 5 ConvLSTM2D layers each with 30 filters each and a kernel size of 3x3. In order to prevent over-fitting, the ConvLSTM2D layers are followed by dropout and batch normalisation layers, and l2 regularisation is employed on the first two ConvLSTM2D layers. At the end of the network, global average pooling is used to connect to an dense output layer of size 3 with a softmax activation function. The network is trained for 30 epochs with a batch size of 200 events per step, where the images and histograms have amplitudes normalised to the range [0,1]. This takes approximately 3 days on a NVidia 1080Ti GPU.

\begin{table}[t]
    \centering
    \begin{tabular}{c|c|c}
    \textbf{Method} &\textbf{Histograms Used} & \textbf{Ordering}\\
    \hline
    A& Timing& Mean Peak Time\\
    B& Charge, Timing, Mean Amplitude, Peak Amplitude & Mean Peak Time\\ &FWHM,RMS, RT, FT& \\
    C& Charge & Size Parameter\\
    D& Charge & Mean Peak Time\\
    \end{tabular}
    \caption{Summary of Methods Used}
\end{table}

Convolutional neural networks are generally designed to handle square images, but most IACT camera designs are not square. In the case of CHEC, the camera plane has holes in order to accommodate the flasher calibration system. This risks creating a geometric bias in the machine learning analysis, where events are not treated equally depending on where they occur in the camera plane. Whilst more sophisticated tools are under development to attack this problem \cite{dl1}, for the purposes of this work we simply crop the CHEC images and histograms to the central region of the camera (32x32 pixels). We do not perform any additional pre-selection cuts or image cleaning.

\section{Results}
\begin{figure}
  \centering
  \includegraphics[width=\textwidth]{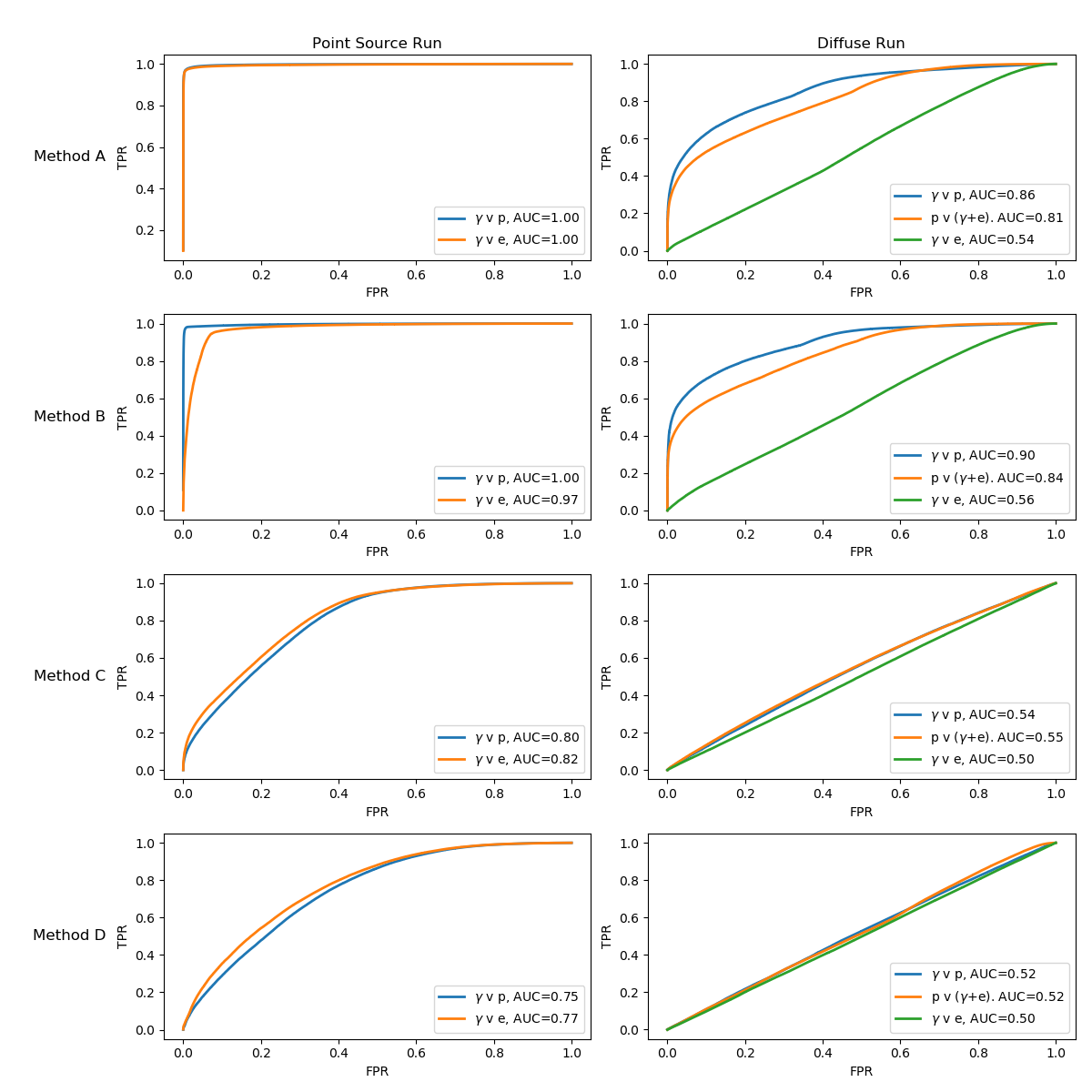}
  \caption{Receiver Operator Characteristic (ROC) curves for the four methods described displaying the True Positive Rate (TPR) against the False Positive Rate (FPR).  The associated Area Under Curve (AUC) metrics are also shown.  In the point source run case, we consider the ROC curves for classification between $\gamma$-rays and protons and between $\gamma$-rays and electrons. This demonstrates the combined classification power of both morphological and directional information. The equivalent curves for the diffuse case highlight solely the morphological classification power. The proton versus ($\gamma$+e) curve demonstrates the difficulty  of distinguishing between $\gamma$-rays and electrons.
  }
  \label{fig:ROC}
\end{figure}

\begin{figure}
\includegraphics[width=0.48\textwidth]{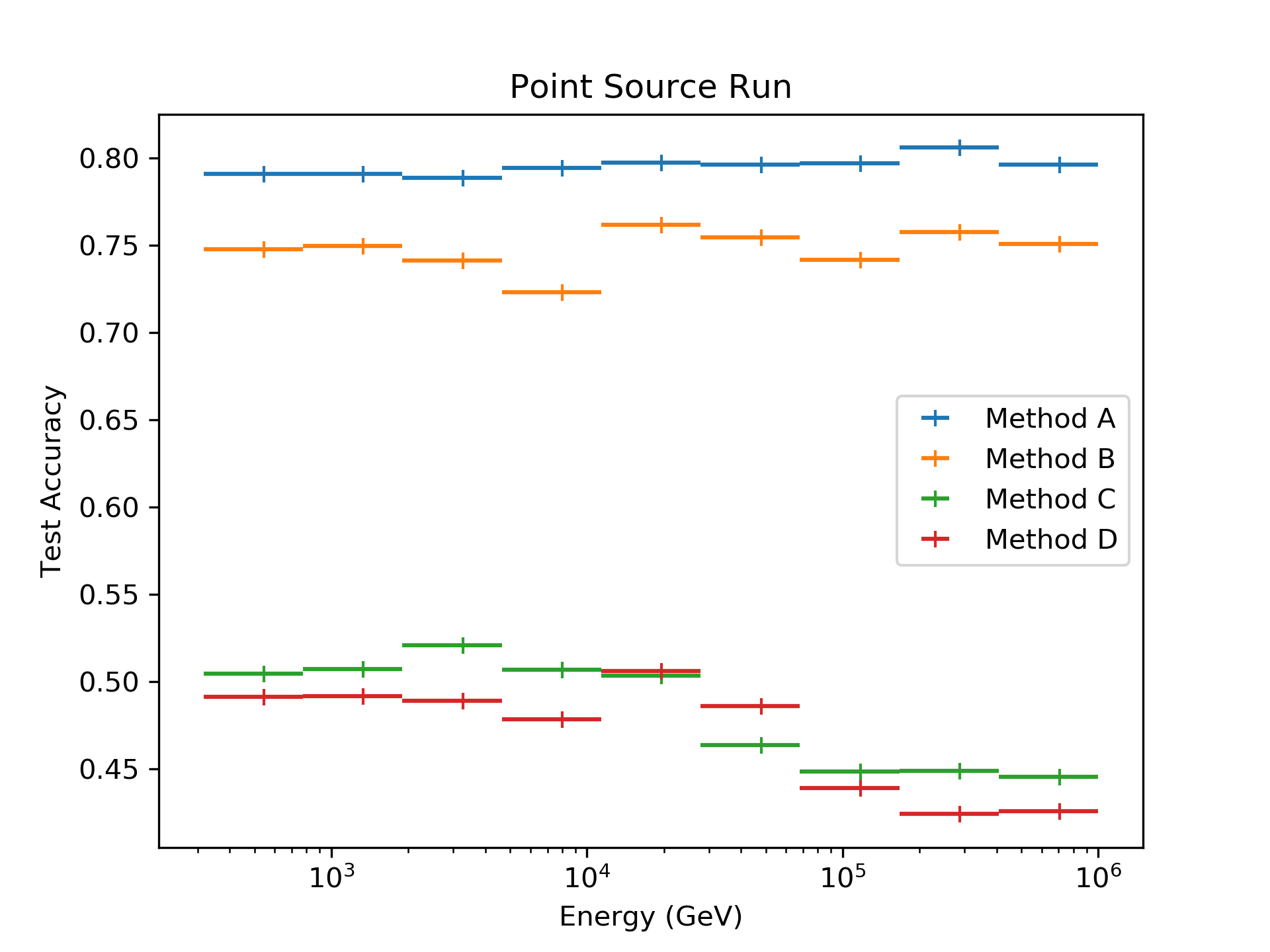}
\includegraphics[width=0.48\textwidth]{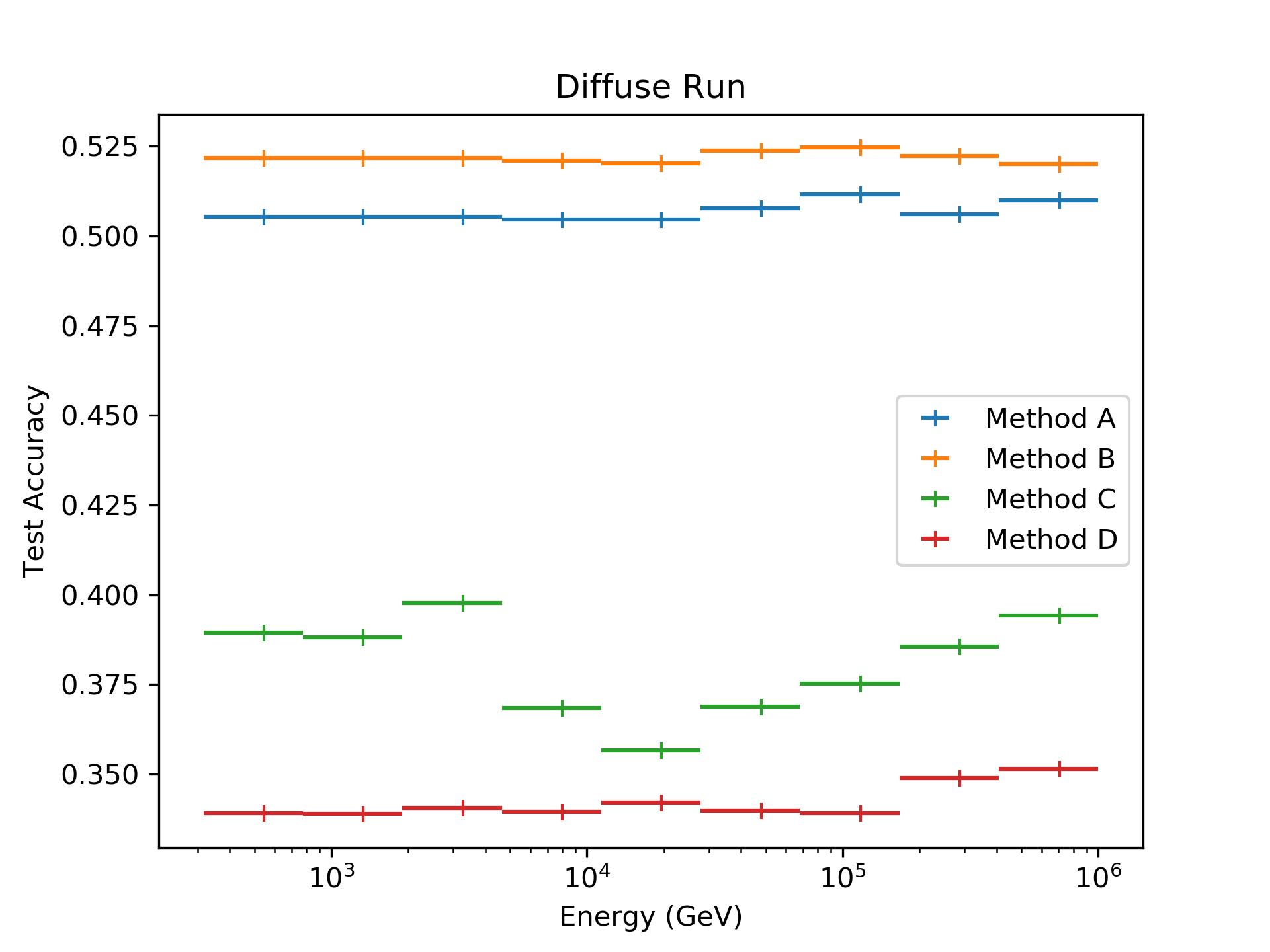}
\caption{Classification categorical accuracy as a function of energy for the point source run (left) and the diffuse run (right). The width of the x error bars depicts the size of the bin.}
\end{figure}

In the case of the point source runs, there is significant confusion between the hadronic and leptonic backgrounds as they are both diffuse, suggesting that the ConvLSTM2D is capable of using information about the shower orientation and position in the image. However, the Method A using the timing histogram significantly outperforms Method C from the current literature. In the case of the diffuse events, for all four methods there is significantly more confusion between electron and gamma-ray events. This suggests that electron induced events will remain a source of irreducible background for IACTs, although Methods A and B still outperform the Shilon technique on a like-for-like basis. Additionally, the classification accuracy curves for methods C and D demonstrate that the ordering of the images in hybrid CRNN methods has some effect.

To evaluate the classification accuracy as a function of energy, we logarithmically bin the test datasets into nine bins. We then take the networks trained on the complete dataset and calculate the categorical test accuracy for each bin and each method. The categorical test accuracies are remarkably constant as a function of energy for all four methods, especially considering that in the lowest energy bin there are O(200,000) events whereas in the highest energy bin there are O(500) events.

\section{Summary and Outlook}
We have investigated four methods of applying new machine learning methods to the task of classifying events for IACTs, finding that the use of timing information is an effective method of event discrimination. 

In considering future prospects, we must stress that in the study presented here, our CRNN algorithms have been trained entirely with simulated data.  There are potentially serious issues to be considered when applying these CRNN based methods with real data: variations in night sky background levels, dead pixels or front end electronics, bright stars in the field of view, cloud cover and the ageing of telescopes could all create biases in the CNN response, and not all of these effects are trivial to simulate. An investigation into how to mitigate these issues will be presented in a later work.

\section{Acknowledgements}
This work used the EU Supercomputing grid. We thank the CTA consortium for allowing us to use CTA simulation models. SS acknowledges an STFC Ph.D. studentship. GC, JW and TA acknowledge support from STFC grants ST/S002618/1 and ST/M00757X/1. GC acknowledges support from Exeter College, Oxford.

\end{document}